\documentclass[sigconf,natbib=true]{acmart}
\usepackage{makecell}
\usepackage{subcaption}
\usepackage{graphicx}
\usepackage{tcolorbox}
\usepackage{courier} 
\usepackage{setspace} 
\usepackage{url}
\AtBeginDocument{%
  }

\begin{document}

\title{\textsc{Eco-Amazon}: Enriching E-commerce Datasets with \\Product Carbon Footprint for Sustainable Recommendations}

\author{Giuseppe Spillo}

\affiliation{%
  \city{University of Bari Aldo Moro}
  \country{Italy}
}
\email{giuseppe.spillo@uniba.it}

\author{Allegra De Filippo}
\affiliation{%
  \city{University of Bologna}
  \country{Italy}
}
\email{allegra.defilippo@unibo.it}

\author{Cataldo Musto}
\affiliation{%
  \city{University of Bari Aldo Moro}
  \country{Italy}
}
\email{cataldo.musto@uniba.it}

\author{Michela Milano}
\affiliation{%
  \city{University of Bologna}
  \country{Italy}
}
\email{michela.milano@unibo.it}

\author{Giovanni Semeraro}
\affiliation{%
  \city{University of Bari Aldo Moro} 
  \country{Italy}
}
\email{giovanni.semeraro@uniba.it}

\renewcommand{\shortauthors}{Spillo et al.}

\begin{abstract}
In the era of responsible and sustainable AI, information retrieval and recommender systems must expand their scope beyond traditional accuracy metrics to incorporate \textit{environmental sustainability}. However, this research line is severely limited by the lack of item-level environmental impact data in standard benchmarks. This paper introduces \textsc{Eco-Amazon}, a novel resource designed to bridge this gap. Our resource consists of an enriched version of three widely used Amazon datasets  (\textit{i.e.,} Home and Kitchen, Clothing, and Electronics) augmented with Product Carbon Footprint (PCF) metadata. $CO_2e$ emission scores were generated using a zero-shot framework that leverages Large Language Models (LLMs) to estimate item-level PCF based on product attributes. Our contribution is three-fold: \textit{(i)} the release of the \textsc{Eco-Amazon} datasets, enriching item metadata with PCF signals; \textit{(ii)} the LLM-based PCF estimation script, which allows researchers to enrich any product catalogue and reproduce our results; \textit{(iii)} a use case demonstrating how PCF estimates can be exploited to promote more sustainable products. By providing these environmental signals, \textsc{Eco-Amazon} enables the community to develop, benchmark, and evaluate the next generation of sustainable retrieval and recommendation models. 
Our resource is available at \url{https://doi.org/10.5281/zenodo.18549130}, while our source code is available at: \url{http://github.com/giuspillo/EcoAmazon/}.

\end{abstract}



\maketitle

\section{Introduction}

The environmental impact of digital services and online consumption has become an increasingly important concern for both researchers and policymakers \cite{he2023modeling}. While e-commerce platforms have improved accessibility and efficiency, they also contribute substantially to global carbon emissions through complex supply chains and manufacturing processes. As online marketplaces continue to scale, supporting more sustainable consumption choices emerges as a critical challenge for Information Retrieval (IR) and recommender systems (RS) research. In this context, Product Carbon Footprint (PCF), commonly expressed as \textit{cradle-to-grave} $CO_2$-equivalent ($CO_2e$) emissions, serves as a key information signal for enabling sustainability-aware retrieval, ranking, and recommendation \cite{zhou2024advancing}.

The most common strategy to quantify PCF relies on the popular Life Cycle Assessment (LCA) methodologies, which provide accurate estimates of product-level emissions \cite{nurdiawati2025recent} by using process-level data, expert knowledge, and proprietary emission databases \cite{koyamparambath2022implementing}.
Unfortunately, while platforms like Environdec\footnote{https://environdec.com/} provide PCF information for thousands of products, these items are only loosely aligned with those appearing in datasets for IR and RS research. As a result, there is a lack of item-level environmental annotations \cite{wasilewski2024sustainability}, and this scarcity of PCF-enriched resources hinders reproducibility, limits comparison across methods, and slows progress on sustainability-aware retrieval and recommendation.

Recent advances in LLMs offer a promising opportunity to address this gap, since they encode broad domain knowledge derived from large-scale textual corpora and have demonstrated the ability to infer latent attributes from unstructured descriptions in \textit{zero-shot} settings \cite{guo2024can,wu2024towards}. This capability has motivated initial efforts to approximate environmental indicators, including carbon emissions, directly from textual data. However, existing studies are typically restricted to narrow domains, small product sets, or proprietary pipelines \cite{luo2023autopcf,li2025pcf}, and rarely result in reusable research artifacts. Consequently, the IR and RS communities still lack shared, well-documented resources exposing PCF information.

This work addresses this limitation by releasing \textsc{Eco-Amazon}, an enriched version of three widely used Amazon datasets (\textit{i.e.,} Electronics, Home and Kitchen, and Clothing) augmented with PCF score. $CO_2e$ emission scores are obtained through a zero-shot prompting framework \cite{kojima2022large} that leverages LLMs to map product characteristics to LCA benchmarks. Our framework exploits standards such as the GHG Protocol\footnote{https://ghgprotocol.org/} and ISO 14040/14044\footnote {https://www.iso.org/standard/37456.html} and relies on publicly available product metadata and free-text descriptions, without requiring domain-specific training or proprietary data.

Along with the datasets, we also released our estimation script, which allows researchers to further enrich product catalogs and reproduce our results. To support this research line, we also release a curated subset of products with \textit{ground-truth PCF values} derived from official Environmental Product Declarations, which can be used to benchmark alternative PCF estimation strategies. 

Finally, to demonstrate the practical utility of the resource, we present a use case in which PCF scores are incorporated into a recommendation pipeline via post-hoc re-ranking \cite{jannach2010recommender}. The results show that including PCF scores can guide users toward lower-impact alternatives with minimal loss in recommendation accuracy. Importantly, this use case demonstrates one of the several applications of the proposed resource, which is designed to support a wide range tasks, including sustainability-aware ranking, diversification, and multi-objective optimization. In summary, this work makes the following contributions: 

\begin{itemize}
    \item we address the scarcity of environmental metadata by releasing publicly available, PCF-enriched e-commerce datasets suitable for IR and RS research;
    \item we release an open source framework for estimating PCF via LLMs using textual data;
\item we confirm the effectiveness of the resource in a use case where PCF values are used as a re-ranking factor in a sustainability-aware recommendation algorithm.
\end{itemize}

By releasing a shared resource for PCF estimation, this paper aims to foster reproducible, sustainability-oriented research and to support work at the intersection of IR, RS, and sustainable AI. 

The rest of the paper is organized as follows. In Section 2, we provide an overview of related work. Next, we describe our methodology for estimating PCF using LLMs and present the resource. Finally, we introduce our case study in sustainable RS, outline the conclusions, and provide directions for future work.

\section{Related Work}

The quantification of the PCF is traditionally governed by the LCA framework. As defined by the European Commission \cite{eulca}, a \textit{"cradle-to-grave"} LCA systematically evaluates five sequential stages: r\textit{aw material extraction, production, distribution, product usage, and end-of-life management}. Methodological standardization for these assessments is provided by international protocols such as the previously mentioned ISO 14040 and the GHG Protocol. 

In the current state of the art, it is common to calculate PCF using databases such as the Inventory of Carbon and Energy (ICE) \cite{hammond2008inventory}, EcoInventDB \cite{wernet2016ecoinvent}, or interfaces such as the Climatiq API\footnote{https://www.climatiq.io/} and the previously mentioned Environdec. However, these tools are hard to use for large-scale estimation of e-commerce catalogs. Indeed, the primary bottleneck lies in their lack of item-level granularity; these repositories do not cover the vast, heterogeneous catalogs characteristic of modern e-commerce. Furthermore, the need for manual mapping between product descriptions and LCA emission factors makes the approaches non-scalable. High licensing costs and the proprietary nature of these databases further restrict their utility for open research. \textbf{Consequently, there is currently no publicly available, large-scale resource that provides item-level environmental metadata for the IR and RS community.}

Accordingly, the research began along the path of automated environmental impact assessment. One of the first attempts was proposed by Algren et al. \cite{algren2021machine}, leading to subsequent applications in industrial carbon accounting \cite{gonzalez2022explainable, wang2023design}. Recent work has increasingly leveraged LLMs to infer environmental indicators from unstructured text \cite{preuss2024large}. These attempts include frameworks such as AutoPCF \cite{luo2023autopcf}, which utilizes GPT-based zero-shot inference and semantic matching to replace expert input, and PCF-RWKV \cite{li2025pcf}, which employs low-rank adaptation for task specialization. However, these models focus on narrow categories (\textit{e.g.}, steel, textiles, batteries) and have only been validated on restricted datasets. \textbf{Most importantly, neither of these approaches has resulted in a released, reusable research artifact that can be integrated into broader retrieval or recommendation pipelines.}

To our knowledge, our previous work \cite{Vicenti25} represents the only attempt to enrich large-scale IR and RS catalogs with PCF information. However, while our previous contribution focused exclusively on the \textit{electronics} domain, this resource significantly broadens the scope by incorporating diverse multi-domain datasets, benchmarking more LLMs, providing an expanded curated ground truth, and offering fully reproducible scripts for estimation. 

To sum up, this work introduces the first resource for assessing LLM-based PCF for e-commerce catalogs at scale, and it is distinguished from existing literature by three key factors:

\begin{enumerate}
\item We provide a generalized, model-agnostic pipeline that utilizes zero-shot inference on public metadata, removing the dependency on closed, proprietary emission databases.
\item Unlike previous studies restricted to small catalogs, we release an enriched dataset based on the Amazon corpus (\textit{i.e,} \textsc{Electronics, Home and }\textsc{Kitchen, Clothing}), representing a massive increase in scale and item diversity.
\item While sustainability in RSs has recently received significant attention \cite{spillo2023towards, spillo2024recsys}, research has largely focused on the energy efficiency of the models themselves. Our resource provides the necessary signals to investigate item-level sustainability, enabling multi-domain, multi-LLM studies of PCF-aware re-ranking in recommendation scenarios.
\end{enumerate}

\section{PCF Estimation through LLMs}

To estimate PCF, we implemented a \textit{zero-shot} prompting strategy that exploits LLMs and bypasses the requirement for domain-specific fine-tuning or few-shot examples. In this section, we introduce our strategy and validate it through both a quantitative and qualitative evaluation.

\subsection{Prompt Design}
The primary objective of our strategy is to bridge the gap between LLMs' latent reasoning capabilities and the official information available in commercial databases. Specifically, when designing the prompts, we force the LLM to adhere to international carbon accounting standards, such as the GHG Protocol and ISO 14040/14044, by incorporating them directly into the prompts as constraints. In this way, the resource ensures that the outputs are not merely statistical approximations but are structured according to recognized LCA principles. 
In particular, the pipeline follows a rigorous two-step process, which is encoded in our prompt (see \textit{Box 1}).
\begin{enumerate}
\item \textbf{Use of Official Data:} first, the system attempts to identify and extract official carbon footprint data from Environmental Product Declarations (EPDs) or manufacturer reports.
\item \textbf{LLM-Driven Inference: }In the absence of such documentation, the framework triggers the LLM to synthesize a PCF estimation derived from unstructured product metadata and descriptions. This happens in the vast majority of e-commerce catalogs.
\end{enumerate}

Such a \textit{two-step} strategy enables \textsc{Eco-Amazon} datasets to include reliable PCF scores while maintaining the breadth required for large-scale IR and RS tasks. By releasing both the PCF estimations and the code used to generate them, we enable a transparent assessment of model performance and error across diverse product categories, as detailed in the subsequent evaluation.
 
\begin{tcolorbox}[colback=gray!5!white, colframe=black, title=Zero-shot Estimation of PCF through LLMs (Box 1), fonttitle=\bfseries, boxrule=0.5pt]
\scriptsize
\ttfamily
You are an expert in life-cycle analysis (LCA) and CO2e emission calculations for electronic products.\\
You must estimate the CO2e emissions, based on the entire life cycle (cradle-to-grave), for the following electronic product.\\
Product data: \{product\_data\}\\

INSTRUCTIONS:\\
1. FIRST, check if there are any official carbon footprint reports or environmental product declarations (EPD)\\ 
\hspace*{1.5em}from the manufacturer for this specific product.\\
\hspace*{1.5em}If found, use these official values as your primary source.\\
2. If NO official manufacturer reports are available, then estimate emissions following these protocols:\\
\hspace*{1.5em}- GHG Protocol Product Standard for system boundaries and calculation methodology\\
\hspace*{1.5em}- ISO 14040/14044 for Life Cycle Assessment principles\\
\hspace*{1.5em}- PAS 2050 and ISO/TS 14067 for carbon footprint calculation guidelines\\
3. For estimation, consider:\\
\hspace*{1.5em}- Main materials composition\\
\hspace*{1.5em}- Manufacturing processes\\
\hspace*{1.5em}- Transportation\\
\hspace*{1.5em}- Use phase energy consumption\\
\hspace*{1.5em}- End-of-life disposal\\
4. Use the most recent emission factors and scientific data available\\
5. Document your sources and assumptions in the explanation\\
6. Clearly state if you're using manufacturer data or estimation\\

Reply ONLY with a JSON object containing these exact fields:\\
\{\{\\
\hspace*{1.5em}"co2e\_kg": <number>,\\
\hspace*{1.5em}"source": <if "manufacturer report" or "estimation">,\\
\hspace*{1.5em}"explanation": "<detailed explanation including data source>"\\
\}\}
\end{tcolorbox}

\subsection{Quantitative Evaluation}

The most straightforward strategy to validate our approach is to calculate the \textit{average error} in the PCF estimates produced by the LLMs. However, as previously explained, most items in e-commerce catalogs lack official data about their PCF, making it hard to assess the effectiveness of our pipeline in the absence of \textit{ground-truth} data. Accordingly, to fulfill this task, we searched for items with Environmental Product Declarations issued by producers, since these items report official information on their PCF.

In particular, we focused on three e-commerce domains relevant to IR and RS scenarios (\textit{i.e.}, \textsc{Electronics, Clothing, Home and Kitchen}). Domains were chosen because they include \textit{physical consumer goods}. Unlike digital products, whose emissions are largely indirect and infrastructure-dependent, physical goods allow meaningful, comparable footprint estimation. In total, we collected PCF information for 159 items (63 for \textsc{Electronics}, 52 for \textsc{Clothing}, and 44 for \textsc{Home and Kitchen}), which we use as the \textit{ground truth} in this experiment. Our resource includes the list of items along with their PCF values to foster further research in this area.

Next, we run the zero-shot prompt from Box 1 on all the items in the list by using \textit{GPT-o3-mini} and \textit{Gemini-2.5-flash} as LLMs, and we calculated: \textit{(i)} Mean Absolute Error (MAE) obtained by comparing the ground-truth $CO_2e$ and LLMs estimation; \textit{(ii)} Spearman rank coefficient, normalized to the $[0,1]$ interval instead of the original $[-1,-1]$ interval for simpler interpretability, and (iii) NDCG calculated by comparing the ground-truth ranking of the items, ordered based on their descending $CO_2e$ emissions, with the ranking obtained based on the estimation calculated by the LLMs. The evaluation metrics were chosen because they offer different perspectives on the approach's effectiveness. Indeed, while 
MAE indicates how precise the estimate is,  NDCG and Spearman help us assess whether our approach correctly \textit{ranks} items from the more carbon-friendly to the less. The results of this comparison are reported in Table \ref{tab:pcf_llm_results}.

\begin{table}[h]
\centering
\resizebox{0.49\textwidth}{!}{
\begin{tabular}{lcccccc}
\toprule
& \multicolumn{3}{c}{\textbf{GPT-o3-mini}} & \multicolumn{3}{c}{\textbf{Gemini-2.5-flash}} \\
\cmidrule(lr){2-4} \cmidrule(lr){5-7}
\textbf{Category} & MAE & Spearman & NDCG & MAE & Spearman & NDCG \\
\midrule
Electronics    & 48.1 & 0.963 & 0.848 & 73.1 & 0.964 & 0.871 \\
Home and Kitchen & 102.91 & 0.919   & 0.978   & 153.22 & 0.939   & 0.988   \\
Clothing       & 7.6 & 0.756   & 0.942   & 8.1 & 0.821   & 0.953   \\
\bottomrule
\end{tabular}
}
\caption{PCF estimation performance. MAE scores are reported in kg. Spearman scores are normalized $\in [0,1]$.}

\label{tab:pcf_llm_results}
\end{table}

\vspace{-10px}

Table 1 reports the performance of \textit{GPT-o3-mini} and \textit{Gemini-2.5-flash} in estimating PCF across the domains. Regarding absolute accuracy, GPT-o3-mini generally outperforms Gemini-2.5-flash, consistently achieving lower MAE across all domains. However, the MAE values remain non-negligible, particularly in the \textsc{Home and Kitchen} category, where both models exhibit their highest errors. These results suggest that while LLMs can approximate the carbon intensity of products, they still face challenges in providing precise, gram-for-gram numerical estimations, likely due to the complexity of LCA data and the lack of standardized reporting across industries.

However, despite the observed margins of error in absolute terms, the ranking performance of both models is remarkably robust. The normalized Spearman correlation and NDCG scores are consistently high across categories, with both values often exceeding 0.9. This indicates that even when the exact PCF value is misestimated, the models effectively capture the relative environmental impact of products, correctly identifying which items are more or less sustainable. This finding is undoubtedly important for the development of sustainable-aware IR and RS solutions. In these applications, accurately ranking greener alternatives is often more critical than absolute numerical precision, as it enables the system to reliably guide users toward more eco-friendly consumption choices.

\begin{figure}
    \centering
    \includegraphics[width=0.4\textwidth]{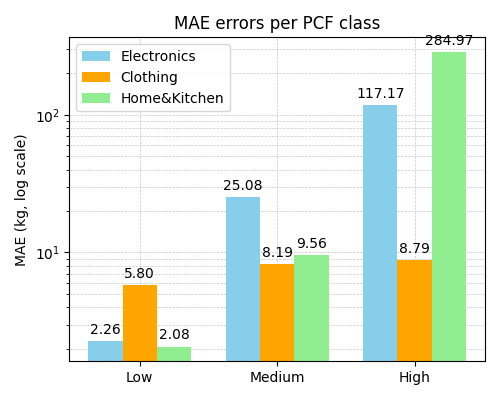}
    \caption{Analysis of MAE per PCF class}
    \label{fig:bucket}
\end{figure}

To provide a more granular understanding of the model's performance, Figure \ref{fig:bucket} breaks down the MAE of the best-performing model (i.e., \textit{GPT-o3-mini}) across three PCF impact categories: \textit{Low, Medium, and High}, obtained by splitting the items in the ground truth based on their percentile. The results reveal that the average error is heavily skewed by high-impact items. While the model achieves high precision for low-impact products—maintaining an MAE below 6 kg $CO_2e$ across all domains, the error increases sharply for high-impact categories, reaching 284.97 in \textsc{Home and Kitchen}. This indicates that the global MAE is disproportionately driven by these carbon-intensive outliers. However, the model remains most accurate in the low-impact tier, which is the most critical segment for identifying truly sustainable alternatives in IR and RS tasks.

Of course, these results can be considered a baseline. By releasing the list of items along with their ground-truth PCF, we aim to foster research in the area and enable comparisons with other, more sophisticated approaches to PCF estimation, \textit{e.g.}, fine-tuning LLMs or Retrieval-Augmented Generation techniques.

\subsection{Qualitative Analysis}

To validate the reliability of PCF estimates, in Table \ref{tab:qualitative_analysis}, we report selected output values from our prompt. In particular, we selected items with different estimated PCF ranges, from low values (just a few kilograms) to higher values. Of course, the complete list of items in the ground truth, together with the estimates returned by GPT and Gemini are included in our repository.

Our analysis reinforces the previous quantitative findings, illustrating that while absolute errors can be substantial, they are often concentrated among high-impact outliers. For low-impact products, the models show a satisfying precision; for example, the estimated PCF for the Logitech Wireless Mouse (4.00 kg $CO_2$) and the 501® Original Jeans (33.40 kg $CO_2$) align almost perfectly with the ground truth. While this high precision can be also due to potential \textit{data leakage} concerns, \textit{i.e.}, data already available in the training of LLMs, we note that this issue is not applicable to our setting: the task aims to estimate an objective, item-level attribute, and therefore any prior knowledge available to the LLM, whether memorized or inferred, does not compromise validity but rather supports the robustness and reliability of the estimated values.

On the other hand, for extreme values such as the Futura Line Toilet Paper, the models exhibit significant numerical deviations. These results confirm that the global MAE is disproportionately driven by carbon-intensive items, whereas the models remain highly reliable for products with a lower environmental footprint. 

\textbf{The most important finding from this qualitative analysis is that the models often maintain a high degree of ordinal consistency when comparing functionally similar items, a feature vital for sustainability-aware recommendation engines.} Within the clothing category, the models correctly preserve the ranking between basic T-shirts and more complex garments, such as wool runners or jeans. A similar trend is evident in the \textit{Home and Kitchen} domain, where GPT correctly identifies the KI Postura+ Chair and the Benchmark Coffee Table as the most sustainable options in their respective subcategories. This ability to accurately rank alternatives ensures that, in the IR and RS context, users are consistently guided toward the most eco-friendly choice. 

\begin{table}[htbp]
    \centering
    \resizebox{0.5\textwidth}{!}{
    \begin{tabular}{lccc}
        \toprule
        \textbf{Item} & \textbf{GT} & \textbf{GPT-o3-mini} & \textbf{Gemini-2.5-flash} \\
        \midrule
        \multicolumn{4}{c}{\textit{Electronics}} \\ \midrule
        Logitech Wireless Mouse & 3.97 & 4.00 & 4.00 \\
        Logitech G PRO Wireless Mouse & 7.84 & 5.25 & 6.35 \\
        Apple TV 3rd gen & 46.00 & 46.75 & 92.50 \\
        Lenovo Chromebook C330 2-in-1 & 263.00 & 203.75 & 178.90 \\
        ThinkPad P1 Gen 5 & 312.00 & 347.50 & 639.25 \\
        Monitor Dell U2717D & 541.00 & 232.50 & 239.32 \\
        \midrule
        \multicolumn{4}{c}{\textit{Clothing}} \\ \midrule
        The Long Sleeve T-Shirt & 4.51 & 6.60 & 6.05 \\
        The T-Shirt & 5.70 & 5.32 & 5.55 \\
        Kids' Multisport Shoes & 6.15 & 6.00 & 7.30 \\
        Men's Wool Runner & 10.36 & 8.17 & 9.50 \\
        501\textregistered{} Original Jeans & 33.40 & 33.40 & 33.00 \\
        \midrule
        \multicolumn{4}{c}{\textit{Home and Kitchen}} \\ \midrule
        Steelcase DO Chair & 82.20 & 135.25 & 78.39 \\
        KI Postura+ Chair & 11.90 & 10.33 & 9.80 \\
        Benchmark OVO Table & 66.90 & 182.50 & 78.07 \\
        Steelcase Lares Desk & 101.00 & 97.00 & 84.76 \\
        Benchmark Coffee Table & 17.10 & 69.75 & 25.48 \\
        Futura Line Toilet Paper & 2040.00 & 1075.00 & 3861.25 \\
        \bottomrule
    \end{tabular}
    }
    \caption{Ground truth (GT) PCF values compared with estimated PCF values.}
\label{tab:qualitative_analysis}
\end{table}

\vspace{-15px}

\section{The \textsc{Eco-Amazon} Datasets}

After the validation of the proposed zero-shot LLM-based PCF estimation strategies, we used our prompt to enrich e-commerce datasets and release our \textsc{Eco-Amazon} resource. 

To this end, we started with the popular Amazon'23 Reviews Data\footnote{https://amazon-reviews-2023.github.io/}. 
For each dataset, information about users' preferences (\textit{i.e.}, ratings and reviews), and items' metadata is available. As previously stated, our resource focuses on three specific domains, \textit{i.e.}, \textsc{Electronics, Clothing, Home and Kitchen}, which are mapped to three datasets of the original resource. 

Before executing our script for PCF estimation, we filtered the datasets using a \textit{k-core filtering}. This is a common choice for work that exploits these datasets for recommendation, as it reduces data sparsity and guarantees robust results \cite{zhou2023comprehensive}.
In particular, we applied a 15-core filtering to all datasets. This reduced the total number of items to about 11.5K for \textsc{Electronics}, 46K for \textsc{Home and Kitchen}, and 60K for \textsc{Clothing}.
Next, to keep computational and inference costs associated with LLM calls lower, we further randomly sampled the last two datasets, resulting in about 21k items for \textsc{Home and Kitchen} and 17K for \textsc{Clothing}. The final statistics are provided in Table \ref{tab:dataset_stats}.

Next, we run our PCF estimation script for all items across all datasets.  The code to run the prompt is available in our repository. As in our previous experiment, we used both \textit{GPT-o3-mini} and \textit{Gemini-2.5-flash} as LLMs. To ensure robustness, each estimation was repeated 4 times, and for each item, we calculated the average emission values across the calls. 
Finally, the PCF estimates obtained by using both LLMs were added to the original dataset and stored alongside the other features (\textit{e.g.}, item descriptions, categories, prices, etc.). \textbf{In total, our resource includes PCF estimations for 49,902 items over the three datasets\footnote{\url{https://doi.org/10.5281/zenodo.18549130}}.} Of course, our pipeline is completely reproducible and extensible, so more items (and even more datasets) can be processed and enriched through our estimation script. This will be done as future work.

\begin{table}[t]
\centering
\small
\resizebox{0.85\columnwidth}{!}{
\begin{tabular}{lccc}
\toprule
\textbf{Dataset} & \textbf{Users} & \textbf{Items} & \textbf{Ratings} \\
\midrule
\textsc{Electronics}    & 21,751 & 11,495 & 464,464 \\
\textsc{Home and Kitchen} & 66,810 & 17,027 & 684,651 \\
\textsc{Clothing}       & 97,608 & 21,380 & 1,070,586 \\
\bottomrule
\end{tabular}
}
\caption{\textsc{Eco-Amazon} Datasets Statistics.}
\label{tab:dataset_stats}
\end{table}

\section{Use Case: PCF-aware Recommender Systems}

Once the estimation phase is completed, each product in the catalogue is enriched with a PCF value. Next, as a use case, we leverage this information in a recommendation scenario, \textit{i.e.}, to promote more environmentally sustainable items within the recommendation list, with the goal of encouraging greener user choices.

\subsection{Description of the Approach}

In our use case, we incorporate PCF information through a post-processing re-ranking strategy, where PCF scores act as a weighting factor applied to the output of a standard recommendation algorithm. Given a user $u$ and an item $i$, we define the sustainability-aware score (SaS) as follows:
\begin{equation}
SaS(u,i)= (1-\alpha) \cdot pred(u,i) + \alpha \cdot PCF_{LLM}(i),
\end{equation}

where $pred(u,i)$ denotes the relevance score predicted by a standard recommendation model, and $PCF_{LLM}(i)$ represents a non-personalized term derived from the estimated PCF of item $i$. This formulation is inspired by post-processing approaches \cite{antikacioglu2017post,balloccu2022post}, which are commonly employed to optimize non-accuracy objectives, such as novelty and diversity \cite{kaminskas2016diversity}. In our setting, the same principle applies: by adjusting $\alpha$, we control the trade-off between recommendation accuracy and environmental sustainability. As $\alpha$ increases, more sustainable items are generally favored.

Overall, our method takes as input the original recommendation list and produces a re-ranked list that integrates both relevance and sustainability signals. In the experimental evaluation, we investigate the extent to which this re-ranking strategy can increase user exposure to environmentally friendly products while preserving recommendation accuracy. To this end, we evaluate our approach using two widely adopted and state-of-the-art recommendation models, \textit{i.e.}, BPR \cite{bpr_model_ref} and LightGCN \cite{lightgcn_model_ref}, selected for their popularity and strong empirical performance. 
First, the models are trained on the available interaction data. Then, we predict all user-item pairs (for unseen items), and apply our re-ranking strategy.
All experiments are conducted using the RecBole framework\footnote{https://recbole.io/} \cite{recbole_paper}.
The inclusion of additional recommendation models is left for future work.
Next, performance is assessed using standard accuracy metrics, including Precision, Recall, F1-score, and NDCG, as well as beyond-accuracy metrics such as the Gini Index (recommendation diversity), average popularity, and tail percentage. Sustainability is evaluated by computing the average $CO_2e$ emissions of the items appearing in each recommendation list. To quantify the impact of the proposed re-ranking, we measure the difference in sustainability between the original and re-ranked recommendation lists. Due to space constraints, we report results only in terms of Recall and the Percentage of Tail Items in the recommendations. However, our repository includes the complete results. Moreover, we release the source code to also evaluate additional models included in RecBole\footnote{\url{https://github.com/giuspillo/EcoAmazon}}.

\subsection{Experimental Results}

\begin{figure*}[htbp]
     \centering
     \begin{subfigure}[b]{0.39\textwidth}\includegraphics[width=\textwidth]{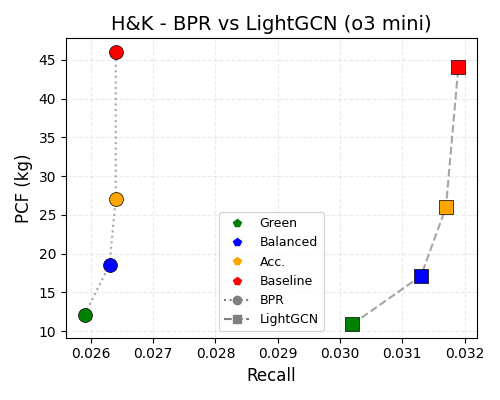}
         \caption{H\&K: Recall vs PCF}
         \label{fig:hk_recall}
     \end{subfigure}
     \hfill
     \begin{subfigure}[b]{0.39\textwidth}
         \includegraphics[width=\textwidth]{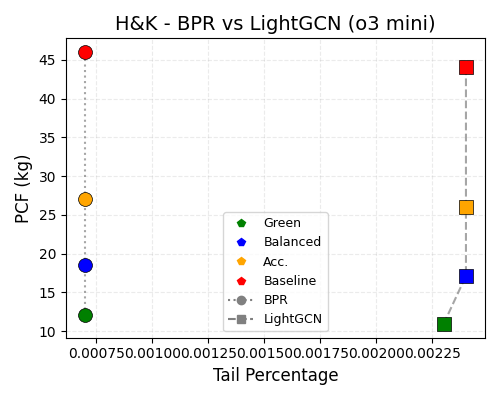}
         \caption{H\&K: Tail Percentage vs PCF}
         \label{fig:hk_tail}
     \end{subfigure}

     \vspace{0.5em} 

     \begin{subfigure}[b]{0.39\textwidth}
         \centering
         \includegraphics[width=\textwidth]{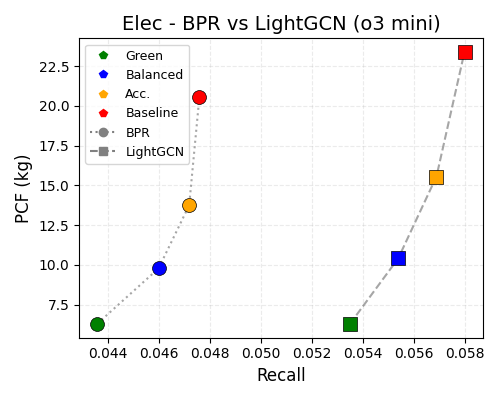}
         \caption{Electronics: Recall vs PCF}
         \label{fig:elec_recall}
     \end{subfigure}
     \hfill
     \begin{subfigure}[b]{0.39\textwidth}
         \centering
         \includegraphics[width=\textwidth]{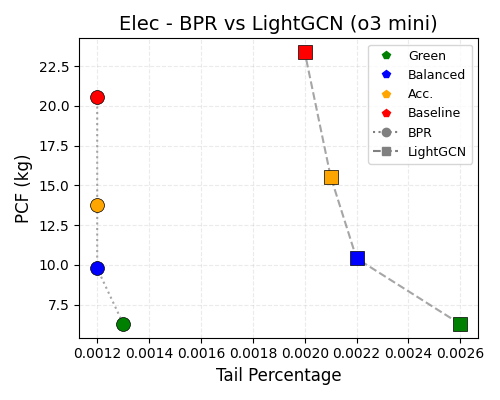}
         \caption{Electronics: Tail Percentage vs PCF}
         \label{fig:elec_tail}
     \end{subfigure}

     \caption{Comparison of Recommendation Models (BPR vs LightGCN) across datasets and metrics, with different $\alpha$ values. $\alpha$ = $0.75$ is the Green focused configuration, $\alpha$ = $0.5$ is the Balanced configuration, $\alpha$ = $0.25$ is the Accuracy focused configuration, and $\alpha$ = $0$ is the baseline configuration, before the post-processing.
     }
     \label{fig:overall_comparison}
\end{figure*}

The results of the experiment are reported in Figure \ref{fig:overall_comparison}. Due to space constraints, we report only the results obtained using the estimates from the best-performing LLM, i.e., \textit{GPT-o3-mini}.  For the same reason, we include results for only two of the three datasets. The complete results are available in our repository.

Overall, the trends highlight the role of $\alpha$ as a control parameter that regulates the importance of \textit{sustainability} in the re-ranking score, with higher $\alpha$ corresponding to lower importance assigned to PCF. 
Accordingly, moving from $\alpha = 0.75$ to $\alpha = 0$ progressively shifts recommendations from environmentally-aware lists to accuracy-oriented ones. 

Across both domains, the results show a consistent trade-off. When more importance is given to PCF ($\alpha$ = 0.75), the average footprint of recommended items is lowest (11–12 kg in Home and Kitchen; 6–7 kg in Electronics, on average), but this comes at the cost of a small reduction in terms of performance. As $\alpha$ decreases, recall improves monotonically and PCF rises markedly (up to 20–23.5 kg). Tail exposure follows a similar, though milder, pattern, suggesting that relaxing PCF constraints not only improves accuracy but also slightly increases long-tail coverage. This is an interesting finding, showing that injecting PCF information also allows for recommending less popular items.

LightGCN consistently achieves higher Recall and tail percentage than BPR. However, in the \textsc{Electronics} domain, this comes at the cost of slightly higher PCF levels. As previously stated, these results are reported for \textit{o3-mini}, with Gemini exhibiting aligned behavior, thus reinforcing the robustness of the observed trade-off across LLMs. From a cross-dataset perspective, \textsc{Electronics}  shows a monotonic reduction in PCF, which comes at the cost of a similar decrease in Recall, whereas \textsc{Home} and \textsc{Kitchen} exhibits a steeper reduction in terms of PCF, which is the ideal behavior. Despite these scale differences, the two datasets show overlapping trends, a pattern also observed in the additional dataset, further supporting the generalization of the findings.

\textbf{Overall, the results confirm the effectiveness of PCF-based re-ranking as a controllable mechanism to navigate the accuracy–sustainability trade-off. Higher $\alpha$ values enable substantial reductions in carbon footprint with limited performance loss, while lower $\alpha$ values recover accuracy as carbon costs rapidly increase—highlighting the importance of intermediate configurations for balanced optimization.}

\section{Discussion and Conclusions}

In this paper, we present the \textsc{Eco-Amazon }datasets. By enriching large-scale, widely adopted Amazon benchmarks with PCF, \textsc{Eco-Amazon} lowers the barrier for developing and evaluating sustainability-aware IR and RS. Our preliminary analyses show that LLM-based PCF estimation, while imperfect at predicting emissions, is sufficiently reliable for ranking-oriented tasks, supporting its use in multi-objective re-ranking scenarios. At the same time, the observed variability across domains and models highlights open challenges in estimation robustness and calibration, motivating future work on hybrid pipelines combining LLM reasoning, structured LCA databases, and retrieval-augmented strategies. 
Overall, we hope this resource will foster reproducible research at the intersection of IR, RS, and sustainable AI, enabling the community to move beyond accuracy-centric evaluation toward multi-objective paradigms that explicitly account for environmental impact.

\bibliographystyle{ACM-Reference-Format}
\bibliography{sample-base}

\end{document}